# Capítulo 14

# Búsqueda de señales de actividad tecnológica en la galaxia

Guillermo A. Lemarchand

**Resumen:** en este trabajo se presenta un análisis de los fundamentos de los programas de búsqueda de señales artificiales de origen extraterrestre en la galaxia, que se han venido desarrollando por más de cinco décadas. Se muestra que el factor determinante para el éxito de estos proyectos de investigación, está dado principalmente por la vida media de una civilización tecnológica. Asumiendo el Principio de Mediocridad, se hacen estimaciones del mínimo número de civilizaciones que pudieran co-existir en la galaxia y de la probabilidad de encontrar alguna señal proveniente de ellas.

**Abstract:** In this article an analysis of the fundamentals used to search for extraterrestrial artificial signals in the galaxy, which have been developing for more than five decades, is presented. It is shown that the key factor for the success of these research projects is given by the technological civilizations lifetimes. Assuming the Principle of Mediocrity, estimations are made to determine the minimum number of civilizations that may co-exist in the galaxy and the probability of detecting a signal from them.

## 1. Introducción

Las primeras propuestas de diseñar metodologías de comunicación con seres de otros mundos comienzan en el siglo XIX (Lemarchand 1992). El 25 de marzo de 1822, el célebre matemático Karl F. Gauss (1777-1855) le envió

Guillermo A. Lemarchand (✉)

Consultor Regional del Programa de Ciencias Básicas e Ingeniería (2008-2010) de la Oficina Regional de Ciencia de la UNESCO para América Latina y el Caribe y Director del Proyecto SETI en el Instituto Argentino de Radioastronomía, C.C. 8 –Sucursal 25, C1425ZAB, Buenos Aires, Argentina

lemar@correo.uba.ar

una carta al astrónomo de Bremen, Wilhelm Olbers (1758-1840), mostrándole que con el uso de aproximadamente 100 espejos planos de un metro cuadrado cada uno, utilizados en conjunto, los humanos podrían llegar a comunicarse con los hipotéticos seres de la Luna. Luego agrega: *"si hiciéramos contacto con nuestros vecinos cósmicos, éste sería un descubrimiento mayor que el de América…"*

En forma independiente, Joseph von Littrow (1781-1840) director del Observatorio Astronómico de Viena, propuso cavar canales con formas de triángulos y círculos de decenas de kilómetros de diámetro, en el desierto de Sahara, para luego verter petróleo y encenderlos de noche. De esta manera, los habitantes del suelo selenita, podrían inferir que las luces con formas geométricas que aparecían en lado oscuro de la Tierra se debían a la presencia de seres inteligentes en su superficie.

En 1869, el inventor francés Charles Cros propone el uso de un sofisticado sistema de espejos parabólicos para enviar señales de luz hacia Marte. Años después, en 1891, una dama francesa devota de los escritos del astrónomo Camile Flammarion (1842-1925), organizó un concurso internacional, ofreciendo 100.000 francos, para el primer científico que hiciera contacto con habitantes de otro planeta o estrella antes de 1901. Marte estaba exceptuado de la lista porque se suponía que era trivial contactar a los marcianos.

En 1899, el famoso inventor Nikola Tesla (1856-1946), construyó en Colorado Springs, una bobina de 23 metros de diámetro, junto a una bola de cobre de un metro, colocadas en una torre de 60 metros de altura, con el objetivo de enviar señales eléctricas a los marcianos… Como resultado, todas las bombitas eléctricas titilaban a 40 km a la redonda de la torre.

En 1919, el premio Nobel de física Guillermo Marconi (1874-1937) afirmaba que las ondas de radio podrían ser utilizadas para establecer comunicaciones con inteligencias de otras estrellas. Consideraba que civilizaciones de planetas más antiguos que el nuestro podrían disponer de conocimientos muy valiosos para la humanidad de la Tierra. Poco más tarde anunciaría que desde su yate *Electra* había detectado señales radioeléctricas de los marcianos.

Albert Einstein (1879-1955), en una entrevista publicada por el periódico *The New York Times* (1920) llegó a afirmar que creía que Marte y otros planetas podrían estar habitados, pero que las señales de Marconi seguramente se debían a perturbaciones atmosféricas o a experimentos de otros sistemas inalámbricos. Luego agrega "Si otras inteligencias estuvieran intentando comunicarse con la Tierra, yo esperaría que utilicen rayos de luz, que son mucho más sencillos de controlar…"

Con los años, los astrónomos se encargarían de desmitificar la existencia de civilizaciones similares a la terrestre en el resto de los planetas del sistema solar. En 1959, dos físicos de la Universidad de Cornell, Giuseppe Cocconi (1914-2008) y Philip Morrison (1915-2005) publican un trabajo en la célebre revista científica *Nature* donde postulan la utilización de la radioatronomía como herramienta para detectar comunicaciones interestelares originadas en otras civilizaciones de la galaxia. En forma independiente, Frank D. Drake, un radioastrónomo del Observatorio Nacional de Radio Astronomía (NRAO) en Green Bank se encontraba poniendo a punto su equipo para buscar este tipo de señales alrededor de dos estrellas cercanas, Tau Ceti y Epsilon Eridani. Hace cincuenta años, este grupo de pioneros comenzaba una tarea que se prolonga hasta nuestros días: detectar las primeras evidencias de que existen otras civilizaciones tecnológicas en nuestra galaxia.

## 2. Fundamentos de la búsqueda radioastronómica de civilizaciones extraterrestres

Por definición, una civilización tecnológica tiene la capacidad técnica para comunicar o generar cualquier tipo de actividad tecnológica que pueda ser detectable a distancias interestelares. Las señales así generadas, pueden ser intencionales (señales transmitidas para darse a conocer) o no intencionales (generadas como resultado de una actividad tecnológica de propósito local). Cualquier señal de contacto entre dos civilizaciones tecnológicas de nuestra galaxia, está restringida por las leyes físicas que gobiernan el universo. Para ser detectable a distancias interestelares, la señal deberá: (1) requerir una cantidad de energía mínima que supere el ruido de fondo galáctico, (2) viajar tan rápido como sea posible, (3) ser fácil de generar, detectar y direccionar, (4) no debería ser absorbida por el medio interestelar, las atmósferas planetarias y sus ionósferas.

Cuando se analizan las restricciones que imponen las leyes de la naturaleza al establecimiento de comunicación entre dos puntos del universo separados por distancias interestelares, únicamente la radiación electromagnética de alguna frecuencia específica puede cumplir con los cuatro requerimientos anteriores.

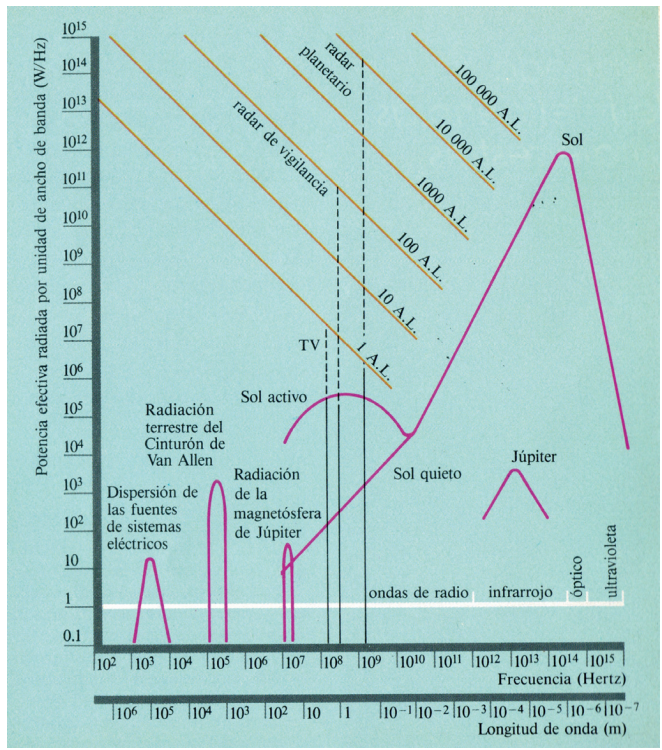

**Fig. 1.** Diagrama indicativo de las potencias efectivas radiadas por diversas fuentes del sistema solar, expresadas en Watts por unidad de ancho de banda en función de la frecuencia y de la longitud de onda. Fuente: Colomb y Lemarchand (1989).

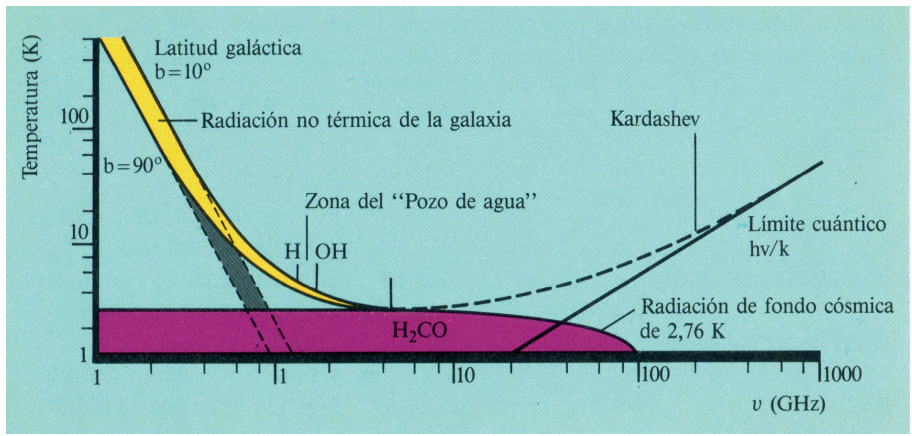

**Fig.2.** Intensidad de ruido de las fuentes naturales de radiación (expresadas en grados Kelvin) en función de la frecuencia. Se representan algunos ejemplos de "frecuencias mágicas" sugeridas para la comunicación interestelar. Fuente: Colomb y Lemarchand (1989).

Para establecer qué región del espectro electromagnético sería la más adecuada para la comunicación interestelar se puede analizar el perfil de radiación de las fuentes naturales y artificiales de nuestro propio sistema solar. El Sol, Júpiter, los radares militares y planetarios, las emisoras de TV y FM, emiten en diferentes frecuencias y con distintas potencias. La Figura 1 presenta el espectro de emisión de las principales fuentes, que son, por otra parte, las que aquí nos interesa considerar. En el eje de las abscisas se representan las frecuencias y se indican las diversas regiones del espectro –ondas de radio, infrarrojo, luz visible, ultravioleta- así como también las longitudes de onda correspondientes a cada frecuencia. En el eje de las ordenadas se representan las potencias efectivas emitidas por unidad de ancho de banda. Las líneas oblicuas muestran las distancias máximas, en años luz, a las cuáles un instrumento de observación "extraterrestre" con una sensibilidad similar a la del Observatorio de Arecibo, estaría en condiciones de captar las emisiones enviadas desde el sistema solar con una frecuencia y potencia determinadas (Colomb y Lemarchand 1989).

Si se pretende establecer contacto con una civilización galáctica, se deberá buscar la manera de que la señal que se emite se haga evidente a la civilización receptora *(Principio Anticriptográfico)*. El análisis de la Figura 1 facilitará comprender qué región del espectro electromagnético y qué potencias deben elegirse para enviar señales artificiales que se distingan del ruido de fondo del sistema solar. Un criterio para reconocer si una señal es natural o artificial es el ancho de banda en el que se concentra la energía de la transmisión. Las fuentes naturales más estrechas que se conocen son los máseres cósmicos, que pueden concentrar la energía de su señal dentro de un ancho de banda de unos cuantos cientos de Hertz. Sin embargo, se pueden generar señales artificiales mucho más estrechas, hasta concentrarlas dentro de anchos de banda de 0,1 Hz que es el límite inferior que impone el centelleo del medio interestelar (Cordes et al. 1997). De esta manera, el principal criterio de artificialidad de una señal que provenga del cosmos estará caracterizado por su alto nivel de monocromaticidad.

La Figura 1 muestra cómo solo, dentro de la región de las microondas, las portadoras de señales de TV, los radares militares y las emisiones de los radares planetarios (Arecibo, Goldstone, Tidbimbila, etc.) tienen potencia suficiente como para ser detectados por otras civilizaciones tecnológicas de la galaxia. Con el objeto de analizar qué parte de esta región del espectro de microondas sería la más adecuada para la comunicación interestelar, en la Figura 2 se representa el espectro de emisión de ruido de fondo galáctico en las frecuencias comprendidas entre 1 y 100 GHz. Además de la radiación de fondo cósmico que queda como remanente del Big Bang, para frecuencias menores a 1 GHz ($10^9$ Hz) existen fuentes de origen galáctico no térmico que son muy intensas. A frecuencias mayores que 100 GHz aparece el ruido cuántico, el fondo irreductible de intensidad creciente con la frecuencia que posee todo sistema receptor. Entonces, la región menos ruidosa para establecer un diálogo interestelar, está comprendida entre 1 y 100 GHz. Ha de observarse que si se quisiera explorar toda esta parte del espectro electromagnético con una resolución de 0,1 Hz, se necesitaría un analizador espectral de $10^{12}$ canales distintos.

Buscando criterios racionales para acotar el problema de esta investigación, a lo largo de los años se fueron proponiendo un conjunto de frecuencias con características especiales, que toda civilización tecnológica que estudiara el universo físico que nos rodea, tarde o temprano, las descubriría. A estas firmas espectrales se las denominó como "frecuencias mágicas" (Lemarchand 1992). La Figura 2 muestra la posición dentro del dial cósmico de muchas de ellas.

Con el avance de las tecnologías de procesamiento de datos, observatorios como el *Arreglo de Radio Telescopios Allen* (ATA), cuentan con la capacidad de analizar simultáneamente todas las frecuencias comprendidas entre 0,4 y 12 GHz, con una resolución espectral del orden del Hz.

La Figura 3 muestra un esquema simplificado de los parámetros a ser considerados, para diseñar una búsqueda radioastronómica de señales artificiales extraterrestres. En lo que sigue, se presentan las relaciones matemáticas a ser utilizadas para determinar las características de transmisión y recepción que deberán reunir los equipos para establecer la comunicación interestelar.

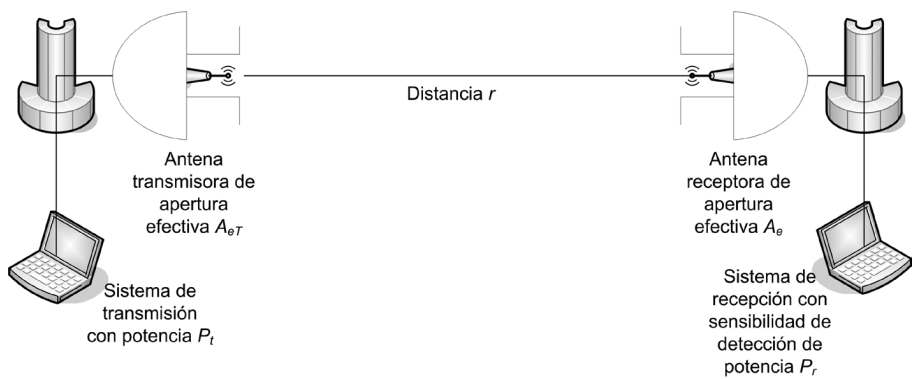

**Fig. 3.** Esquema que representa los principales parámetros a tener en cuenta para la comunicación interestelar. Fuente: esquema del autor.

Asumiendo que una hipotética civilización extraterrestre emplea un radio transmisor que emite isotrópica y homogéneamente señales al espacio, con una potencia $P_t$ dentro de un ancho de banda $B_t$, entonces generaría a una distancia $r$ una densidad de flujo $P_t/B_t 4\pi r^2$. Una antena receptora de apertura efectiva $A_e$ a una distancia $r$ podría detectar las señales con un nivel de potencia en Watts [W] equivalente a:

$$P_r = \frac{B_r P_t A_e}{B_t 4\pi r^2}$$

donde: $B_r$ = resolución espectral del receptor [Hz]
$P_t$ = potencia de transmisión [W]
$A_e$ = apertura efectiva de la antena receptora [m$^2$]
$B_t$ = ancho de banda de la señal transmitida [Hz]
$r$ = distancia entre el transmisor y receptor, [m]

Aquí se asume que $B_r \leq B_t$. Si se considera que la antena transmisora tiene una directividad $D$, la potencia recibida se transforma entonces en:

$$P_t = D \frac{B_r P_t A_e}{B_t 4\pi r^2} = \frac{4\pi A_e}{\lambda^2} \frac{P_t A_e B_r}{4\pi r^2 B_t} = \frac{P_t A_e A_{eT} B_r}{r^2 \lambda^2 B_t} \quad [W]$$

Donde $\lambda$ es la longitud de onda expresada en metros y $A_{eT}$ es la apertura de la antena transmisora en metros cuadrados.

En toda comunicación es importante considerar el cociente entre las potencias de las señales y las del ruido de fondo (SNR). Teniendo en cuenta que la potencia del ruido de fondo es descripta por la relación de Nyquist:

$$P_n = k\, T_{sis} B_r \quad [W]$$

donde: $k$ es la constante de Boltzmann = 1,38 x 10$^{-23}$ [J K$^{-1}$],
$T_{sis}$ = es la temperatura del sistema receptor [K] y
$B_r$ = la resolución espectral del receptor.

Si se hace el cociente entre $P_r/P_n$ se obtiene la siguiente expresión adimensional:

$$\frac{P_r}{P_n} = \frac{P_t A_{eT} A_e B_r}{k T_{sis} B_r r^2 \lambda^2 B_t}$$

La potencia del ruido disminuye aumentando el tiempo de integración *t* en una proporción *($B_r$ x $t$)$^{0,5}$*. De esta manera, la relación señal/ruido (SNR) se transforma en:

$$SNR = \frac{P_t}{P_n/\sqrt{B_r t}} = \frac{P_t A_{eT} A_e \sqrt{B_r t}}{k T_{sis} r^2 \lambda^2 B_t}$$

Si se asume que $B_t = B_r$ (ancho de banda de transmisión igual a la resolución en frecuencia de la recepción) y que $B_r t = 1$ (la duración del pulso se corresponde con la resolución en frecuencia) es posible representar la relación SNR como:

$$SNR = \frac{P_t A_{eT} A_e}{k T_{sis} r^2 \lambda^2 B_r}$$

De allí se puede despejar *r* (distancia entre transmisor y receptor) como:

$$r = \sqrt{\frac{P_t A_{eT} A_e}{SNR \, k T_{sis} \lambda^2 B_r}} \quad [m]$$

Aquí se está asumiendo que hay coincidencia entre los sistemas trasmisores y receptores en la dirección del haz de ambas antenas, frecuencia, anchos de banda y polarización de las ondas electromagnéticas utilizadas.

## Recuadro: Un ejemplo de cómo calcular los parámetros para la comunicación interestelar

*Ejemplo:*
Asumamos que una hipotética civilización extraterrestre se encuentra transmitiendo $10^6$ Wm$^{-2}$, en forma de pulsos de 10 segundos de duración, utilizando ondas polarizadas circularmente hacia la derecha, empleando una antena de 100 m de diámetro y una frecuencia ν = 5 GHz. Además se considera un sistema receptor terrestre que cuenta, también, con una antena de 100 m de diámetro, un receptor que tiene una $T_{sis}$ = 10 K y un analizador espectral que dispone de una resolución máxima de $B_r$ = 0,1 Hz. Para facilitar los cálculos se considera que tanto la antena transmisora como la receptora, tienen un 50% de eficiencia en sus respectivas aperturas y que el sistema receptor es capaz de reconocer la señal con un SNR=3. Determinar a qué distancia se podría encontrar la civilización extraterrestre para que el sistema local la detecte.

*Solución:*
Las aperturas efectivas de las antenas estarán dadas por

$$A_e = A_{eT} = 0,5\pi \frac{d^2}{4} = 4000 \,[\text{m}^2] \quad (d = \text{diámetro de la antena})$$

La longitud de onda estará dada como:

$$\lambda = (c/\nu) = (2,997924562 \times 10^8 \text{ m s}^{-1} / 5 \cdot 10^9 \text{ s}^{-1}) = 0,05995849 \,[\text{m}]$$

Entonces:

$$r = \sqrt{\frac{10^6 \times 4000^2}{3 \times 1,38 \times 10^{-23} \times (0,05995849)^2 \times 0,1}} \cong 10^{19} \,[\text{m}]$$

Considerando que 1 año luz es aproximadamente $10^{16}$ m, se deduce que la distancia a la cual estas dos civilizaciones podrían comunicarse estaría dada por

$$r \cong \frac{10^{19}}{10^{16}} = 1000 \,[\text{años luz}]$$

Para facilitar la discusión se debe profundizar acerca de las implicancias de los valores encontrados en el ejemplo del recuadro. Se puede asumir con bastante precisión, que la densidad de estrellas en un radio de 1000 años luz alrededor del Sol es aproximadamente uniforme, y tiene un valor de 0,01 estrellas por año-luz cúbico. De aquí se infiere que dentro de un volumen esférico de radio 1000 AL, se pueden llegar a encontrar aproximadamente unas $N_e$ = 4 x $10^7$ estrellas. Para las antenas del ejemplo:

$$D = \frac{4\pi A_e}{\lambda^2} = \frac{4\pi \times 4000}{(0{,}05995849)^2} \cong 1{,}4 \times 10^7$$

Por lo tanto, el número máximo de objetos que la antena podría observar es de $\tilde{N} = D = 1{,}4$ x $10^7$. Entonces, como se asumió una densidad constante, en cualquier dirección del cielo una antena de 100 m, siempre tendrá dentro de su haz la siguiente cantidad aproximada de estrellas:

$$\frac{N_e}{D} = \frac{4 \times 10^7}{1{,}4 \times 10^7} \cong 3$$

Aumentando la capacidad de procesamiento de la información recogida, se podría identificar señales artificiales con una relación señal-ruido de SNR = 1, esto implicaría que el rango de distancia desde la cual se detectaría la señal crecería un factor $3^{1/2}$. De esta manera, la cantidad total de estrellas a ser observadas se incrementaría con el cubo de la cantidad anterior, de lo que se deduce que el número de estrellas dentro del haz de la antena en toda dirección sería de aproximadamente 16 estrellas.

Por otra parte, si la densidad de estrellas o la intensidad de la transmisión fuesen menores, el número de estrellas dentro de cada haz de la antena receptora sería substancialmente menor. En este caso, la estrategia de apuntar solo a estrellas cercanas en direcciones fijas, sería mucho más eficiente que la de utilizar la misma antena para hacer un relevamiento de todo el cielo.

## 3.    Las dimensiones del pajar cósmico

Los primeros trabajos detallados en donde se analiza el papel de los distintos parámetros libres involucrados en el diseño de una estrategia de observación comprehensiva, comienzan con el estudio de la NASA conocido como *Proyecto Cíclope*, que data de principios de la década del setenta (Billingham y Oliver, 1971). Un década más tarde, se desarrolló un segundo estudio de este tipo, también dentro del ámbito de la NASA (Drake et al. 1983). Finalmente, en el 2002, el *SETI Institute* desarrolló un trabajo donde establece una estrategia de desarrollo tecnológico para la detección de señales artificiales extraterrestres hasta el 2020 (Ekers et al 2002).

Asumiendo que las ondas electromagnéticas de una determinada frecuencia se constituyen en el canal de comunicación óptimo entre dos civilizaciones galácticas, que no tienen comunicación previa entre sí, se puede presentar las diversas variables desconocidas dentro de un espacio de configuración de doce dimensiones distintas (Lemarchand 1994a). La Tabla 1 presenta una síntesis del mismo. El diseño de cualquier mensaje interestelar deberá contemplar cada una de estas dimensiones y la estrategia de búsqueda, a su vez, deberá estar adecuada a las restricciones que imponen las incertezas de cada dimensión y las limitaciones del estado del arte de las tecnologías disponibles.

Con el advenimiento de los láseres de potencia que pueden concentrar altos niveles de energía en fracciones temporales del nanosegundo, se abrieron nuevos caminos teóricos y observaciones para el desarrollo de la búsqueda de señales artificiales de origen extraterrestre en el rango óptico del espectro electromagnético (Lemarchand et al. 1993 y Lemarchand 1997). Se han desarrollado técnicas especiales para detectar este tipo de señales y procesar los $10^{10}$ bits de información que llegan en cada segundo de observación. Algunas de ellas fueron empleadas en la observación de estrellas cercanas (Howard et al 2004). Por otra parte, liderado por Paul Horowitz de la Universidad de Harvard, se está desarrollando desde 2004, un programa de búsqueda en todo el cielo de pulsos ópticos de nanosegundos de duración (Howard et al 2003).

**Tabla 1.** Dimensiones del espacio de configuración de las variables involucradas en la detección de señales electromagnéticas provenientes de otra civilización. Fuente: Elaboración del autor basado en Lemarchand (1994a).

| Tipo de Dimensión | Propiedades y características de las señales | Restricciones |
|---|---|---|
| Dimensiones de información | $\Delta \nu$ = cubrimiento en frecuencia | Los límites tecnológicos presentes permiten relevar simultáneamente todas las frecuencias entre 0,4 y 12 GHz (por ejemplo ATA) |
| | $\Delta B$ = resolución espectral | Los límites impuestos por el fenómeno de centelleo interestelar, restringen las señales a anchos mínimos de 0.05 a 1 Hz |
| | $p$ = polarización | No existen restricciones tecnológicas para contemplar simultáneamente todas las polarizaciones posibles. |
| | $M$ = tipo de modulación | Se pueden analizar señales continuas, señales que derivan en frecuencia, pulsos, etc. |
| | $\tau$ = tasa de transmisión de la información | *Media* $\rightarrow$ Radio SETI ~$10^{-3}$ s<br>*Alta* $\rightarrow$ SETI óptico ~$10^{-8}$ s |
| | $T$ = ciclo de trabajo de transmisión | Muy importante para la validación de los protocolos de detección. Se necesitan ciclos largos para permitir la verificación de la detección por observatorios independientes. Ciclos cortos de transmisión serían muy difíciles de corroborar. |
| Dimensiones espaciales | $\alpha$ = coordenadas de ascensión recta | Se siguen dos estrategias: relevamientos totales del cielo y observación de estrellas fijas |
| | $\delta$ = coordenadas de declinación | |
| | $r$ = distancia radial y sus dimensiones asociadas:<br>• $P_t$ = potencia de transmisión<br>• $P_r$ = sensibilidad del equipo<br>• $A_e$ = área colectora<br>• SNR | Desde ~ 40 años, Arecibo mantiene la máxima sensibilidad para búsquedas de radio SETI<br>Desde ~ 10 años, se dispone de telescopios ópticos de 10-m pero solo se han usado telescopios de 2-m para SETI óptico<br>*Futuro*: SKA – OSETI espacial |
| | $t$ = sincronización | Se debe apuntar el receptor a una dirección del cielo en el exacto momento en que la señal esté pasando por la Tierra. Se debe sincronizar el momento adecuado con el lugar adecuado en el cielo.<br>*Futuro*: antenas omni-direccionales |
| Dimensiones criptográficas | C = código | Se asume el *"Principio anti criptográfico"* |
| | S = semántica | Es muy difícil de estimar *a priori*. Como contra-ejemplo: asumiendo que realmente fue escrito con un código, aun no se pudo decodificar el *"Manuscrito Voynich"* atribuido a Roger Bacon (1214-1292) |

Tarter (2001) presenta una revisión de las características de los distintos programas observaciones de SETI en el mundo. Desde la Argentina se realizó por primera vez un relevamiento total del cielo del hemisferio sur, utilizando una de las dos antenas de 30 m del Instituto Argentino de Radioastronomía (ver Figura 4). El sistema utilizaba un analizador espectral de 8,4 millones de canales de 0,05 Hz de resolución espectral. Las observaciones se realizaron alrededor de distintas frecuencias mágicas, con resultados negativos (Lemarchand 1998a).

Durante los últimos cincuenta años, se han propuesto distintas estrategias alternativas para justificar ingeniosamente –haciendo uso del conocimiento de las leyes que regulan la naturaleza- distintas restricciones para reducir el número de parámetros libres. Se han propuesto diversas frecuencias mágicas (Lemarchand 1992, 1998b), restricciones a los anchos de banda posibles en el medio interestelar (Cordes et al. 1997), distintas estrategias observacionales (Tarter 2001), formas de sincronizar las transmisiones y recepciones de las señales (Lemarchand 1994b), nuevas alternativas en el procesamiento de los datos (Ekers et al. 2002) y recientemente se ha comenzado a discutir aspectos acerca de la universalidad de los mapas cognitivos para ser utilizados en el diseño de la codificación y semántica de los mensajes interestelares (Lemarchand y Lomberg, 2009).

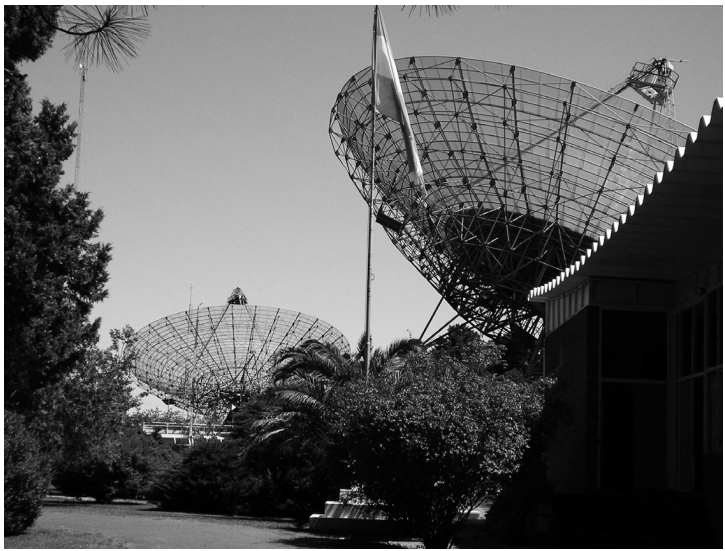

**Fig. 4.** El Instituto Argentino de Radioastronomía (IAR) dependiente del Conicet, dispone de dos antenas de 30 m y puede operar en diversas frecuencias entre 1,4 y 5 GHz. La antena 2 fue utilizada regularmente para hacer el relevamiento de todo el cielo del hemisferio sur, buscado señales de origen artificial con una resolución espectral de 0,05 Hz y una sensibilidad de $10^{-24}$ Wm$^{-2}$. Foto: G.A. Lemarchand.

**Tabla 2.** listado de distintas propuestas realizadas en la literatura científica para encontrar evidencias de actividades tecnológicas en la galaxia. Fuente: Lemarchand (2000a) en donde se encuentra el listado completo de las referencias (*) originales, otros ejemplos se pueden encontrar en Lemarchand (1994b).

| Sistema portador de información | Efecto Observacional | Referencias (*) |
|---|---|---|
| Ondas de radio | Intercepción de comunicaciones entre naves espaciales interestelares | Vallée y Simard-Normandin (1981) |
| Radiación infrarroja | Detección de "Esferas de Dyson" en 10 μm | Dyson (1959, 1966) Jogaku y Nishimura (2000) Slysh (1985) |
| Radiación óptica | Descubrimiento de elementos exóticos en las líneas espectrales de las estrellas (por ej. el tecnecio) | Drake (1964) |
| | Descubrimiento de tecnecio, plutonio, praseodimio, etc. como consecuencias de actividades artificiales de nucleosíntesis, para cambiar las propiedades espectrales de la estrella. Por ejemplo al usar a la estrella como repositorio de los residuos generados por la actividad nuclear en el planeta | Whitmire y Wright (1980) |
| Rayos X | Si se usara una explosión simultánea en el espacio con todo el arsenal nuclear disponible en la Tierra, la misma generaría un pulso omnidireccional de rayos X que sería detectable a 200 años luz de distancia. | Elliot (1973) Fabian (1977) |
| | Secuencias de flashes de rayos X como consecuencia de arrojar material sobre las estrellas de neutrones. | Corbet (1997) |
| | Uso de binarias de rayos X como señales de alerta | |
| Rayos γ | Observaciones astronómicas de radiación γ que presente comportamiento anómalo y que se lo interprete como resultado de la aniquilación de materia y antimateria para su uso como propulsor de sistemas de navegación interestelar | Harris (1986, 1991) Viewing et al (1977) Zubrin (1996) |
| Neutrinos | Detección de haces artificiales de neutrinos usados para la comunicación interestelar | Learned et al (1994) Pasachoff y Kutner (1979) Saenz et al (1977) Subotowicz (1979) Überall et al (1979) |
| Intercambio de Materia | Utilización de codificación de mensajes en sustancias que puedan almacenar información dentro de un código genético de virus o bacterias que puedan sobrevivir por millones de años a los viajes interestelares | Nakamura (1986) Yokoo y Oshima (1979) |
| Técnicas directas | Pequeñas naves sondas espaciales interestelares que generen ecos de nuestras transmisiones de radio | Bracewell (1960, 1975) |
| | Búsqueda de objetos artificiales de origen extraterrestre dentro del sistema solar. | Freitas y Valdes (1980, 1985) |
| | Objetos de origen extraterrestre camuflados en el cinturón de asteroides | Papagiannis (1978, 1985) |
| | Máquinas de von Neumann | Tipler (1980) |
| Exótica | Viajes por agujeros de gusanos, teleportación de estados cuánticos, utilización de leyes de la naturaleza aún desconocidas | Literatura de ciencia ficción |

Cuando se consideran todas las dimensiones posibles del espacio de configuración presentado en la Tabla 1, y se hacen las simplificaciones correspondientes al estado del arte de la tecnología observacional, se llega a la conclusión que las dimensiones a explorar forman un verdadero "pajar cósmico" de un tamaño equivalente a $10^{29}$ celdas distintas. Haciendo una analogía entre la búsqueda de señales de origen extraterrestre y la búsqueda de una aguja de coser en un pajar, fácilmente se podría corroborar que el problema de la búsqueda de evidencias de señales artificiales extraterrestres, es equivalente al problema de buscar una aguja de coser en un pajar, dentro de un volumen 35 veces el del planeta Tierra (Lemarchand 1998b). En cincuenta años de búsqueda, solo se ha explorado $10^{-14}$ veces ese espacio de configuración. Esto implica que estamos aún muy lejos de refutar o corroborar si estamos solos en el universo. La ausencia de evidencia no es evidencia de la ausencia.

Lemarchand (1992, 1994b) hizo una revisión de todas las metodologías alternativas, propuestas en la literatura científica, destinadas a encontrar evidencias de actividades tecnológicas en la galaxia. La siguiente Tabla 2 da cuenta de algunos ejemplos de ellas (Lemarchand 2000a). En todos estos análisis se ha descartado la posibilidad de viajes interestelares tripulados, merced a las grandes limitaciones, en términos de consumo de energía, que tales viajes implicarían (Lemarchand 1992, 2007a).

Posiblemente el primer descubrimiento de una evidencia de actividad tecnológica de origen extraterrestre, se manifieste mediante la detección astronómica de un fenómeno totalmente anómalo e inesperado, cuya única interpretación posible sea adjudicándole un origen artificial.

## 4. La vida media de las civilizaciones tecnológicas

Como se mostró en la Sección 2, el factor clave para el éxito del programa de investigación sobre búsqueda de inteligencia extraterrestre (SETI) es la densidad de civilizaciones tecnológicas en la galaxia. Una densidad baja implica un fracaso prácticamente asegurado de los programas de búsqueda, mientras que una densidad alta aumenta considerablemente una posible detección utilizando la tecnología disponible en el presente.

En 1961, Frank Drake (Pearman 1963) propone una ecuación para estimar el número de civilizaciones tecnológicas en nuestra galaxia. A lo largo de los años se han asignado diferentes valores a los distintos factores de dicha *Ecuación de Drake*, en los cuáles se mostró que el número de civilizaciones tecnológicas en al galaxia (N) es fuertemente dependiente de su último factor *L*, o la vida media de una civilización tecnológica en años (Kreifeldt 1973, Oliver 1975).

El *Principio de Mediocridad* propone que la Tierra y la vida en ella no ocupan ningún lugar especial en el universo. Nuestro sistema planetario, la vida en la Tierra y nuestra civilización tecnológica sería una consecuencia lógica de la evolución biológica en un ambiente adecuado, que resulta de un proceso de miles de millones de años de evolución y selección natural (Hoerner, 1961). En otras palabras, cualquier cosa particular para nosotros tendrá, probablemente, un valor promedio en comparación con otros lugares del universo.

Varios estudios biológicos y ecológicos de muy largo plazo, han demostrado que las diferentes especies en la Tierra, surgen, se desarrollan y se extinguen con similares patrones evolutivos (Charnov 1993, Gurney y Nisbet 1998). La especie humana no puede ser una excepción. El *Homo sapiens* ha roto la "ley" ecológica que establece que los animales grandes, depredadores son raros. En particular, dos innovaciones fundamentales, han permitido a nuestra especie alterar las reglas de la vida en el planeta y permitir así una expansión sin precedentes de la especie: el habla (lo que implica la transmisión instantánea de un rango de composición abierta de pensamientos conscientes) y la agricultura (que hace el mundo pueda producir alimentos en niveles muy superiores a los naturales). Sin embargo, la selección natural no ha dotado al ser humano de un sentido de auto-preservación a largo plazo.

En base a trabajos anteriores (Lemarchand 2000b, 2004, 2006, 2009), aquí se mostrará que los seres humanos se enfrentan a un nuevo tipo de macrotransición: la tecnológica. Lemarchand (2004) definió la *Edad de Adolescencia Tecnológica* (TAA) como la etapa en la que una especie inteligente tiene la capacidad de auto-aniquilase totalmente debido a: (1) el uso de las tecnologías con objetivos auto-destructivos (por ejemplo, la guerra mundial, el terrorismo, etc.), (2) la degradación del medio ambiente del planeta de origen (por ejemplo, el cambio climático, la sobrepoblación, el aumento de las tasas de extinción de especies, etc.), o simplemente (3) por la mala distribución de los recursos físicos, educativos y económicos (diferencia en el nivel de desarrollo entre las sociedades desarrolladas y en camino a desarrollarse). Este último factor podría causar el colapso de la civilización debido a las tensiones generadas por las desigualdades entre las diferentes fracciones de la sociedad global.

Aquí se presenta un enfoque semi-empírico para estimar el periodo de tiempo que se puede extender esta nueva macro-transición societal. Un estudio sistemático de los indicadores pertinentes de comportamiento a largo plazo de nuestra civilización tecnológica es útil, no sólo para hacer una estimación del posible valor de $L$, sino también para identificar cuáles son las variables que necesitan más atención para mejorar y cambiar, a fin de optimizar la expectativa de vida media de la civilización terrestre actual. La comprensión de la dinámica de evolución de estos patrones sociales de largo plazo nos puede ayudar a diseñar diferentes estrategias para evitar la auto-aniquilación.

En un sentido amplio, asumiendo la universalidad del Principio de Mediocridad, la TAA podría ser considerada como el cuello de botella en la evolución de cualquier civilización tecnológica en la galaxia. Una fase en que la continuidad misma de la especie estará obligada a superar la tentación de extinguirse por el mal uso de sus fuerzas acrecentadas.

Aquí se proponen las siguientes tesis: (1) La "etapa de adolescencia tecnológica" debería ser considerada como una transición de fase societal, similar a la ocurrida luego de la invención de la agricultura y el establecimiento de las ciudades (origen de las civilizaciones), (2) Esta "transición" podría ser corroborada o refutada analizando un conjunto de indicadores societales de muy largo plazo, (3) El análisis del comportamiento de la evolución temporal de los distintos indicadores contemplados, podría ser utilizado para estimar la duración de la etapa de adolescencia tecnológica, y (4) El lapso de transición de la adolescencia tecnológica establecería un umbral mínimo de vida media de una civilización tecnológica.

En esta sección se presentan, a modo de ejemplo, algunos de los resultados obtenidos en los trabajos anteriores (Lemarchand, 2004, 2006, 2009).

Se analiza la evolución temporal de una serie de indicadores sociales para entender las constantes de tiempo en las que las sociedades se organizan con el fin de producir cambios en un nivel de macro-comportamiento. Desde la invención de armas de destrucción masiva (por ejemplo, armas nucleares, químicas, biológicas, etc.) por primera vez en la historia evolutiva, la humanidad cuenta con la tecnología como para causar la extinción total de la especie. La enorme inversión en las tareas de investigación científica, desarrollo tecnológico e innovación, con aplicaciones militares y de defensa (aproximadamente 150 mil millones de dólares anuales en tareas de I+D militar) ha generado un crecimiento exponencial en el coeficiente de letalidad del armamento (Figura 5).

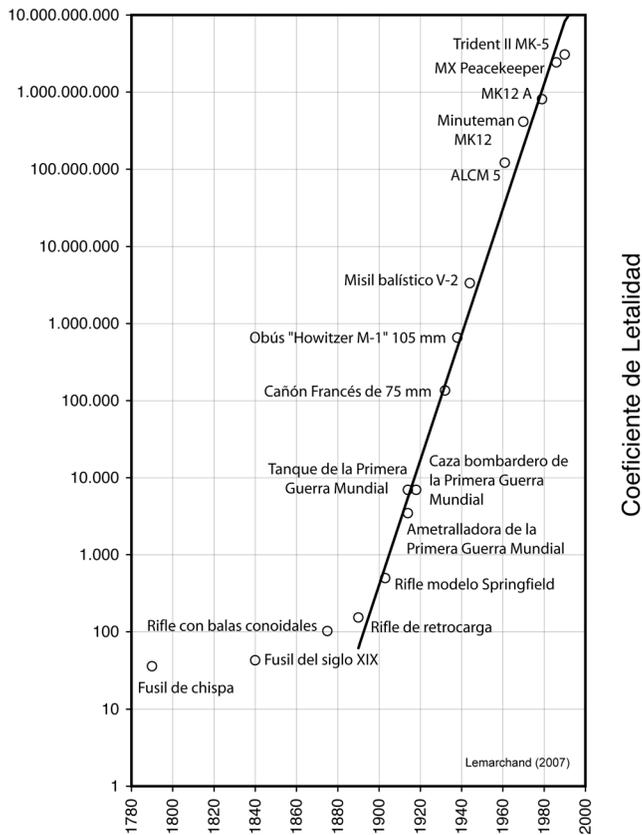

**Fig. 5.** Evolución del coeficiente de letalidad del armamento disponible en función del tiempo. El gráfico está en escala semilogarítmica. El eje de las ordenadas mide número máximo de muertes que un dado armamento genera después de una hora de uso. Fuente: Lemarchand (2007b).

El coeficiente de letalidad mide el máximo número de víctimas por la hora que un arma puede generar, teniendo en cuenta: la velocidad de disparo, el número de objetivos, la eficacia relativa, los efectos de alcance, precisión, fiabilidad, etc. La Figura 5 muestra que la tasa de crecimiento del coeficiente de letalidad ha aumentado en un factor de 60 millones durante los últimos 100 años o 600.000 por año. Este hecho implica que desde hace unas décadas la humanidad dispone ya de capacidad tecnológica para aniquilar completamente nuestra civilización en un corto período de tiempo.

La Figura 6 muestra la distribución del armamento nuclear mundial desde 1945 al presente. En el año 1986 se llegó a tener 70.586 ojivas nucleares desplegadas en el mundo. En la actualidad (veinte años después de la caída del Muro de Berlín) todavía existen 23.360 ojivas emplazadas, que equivalen a casi once veces el umbral mínimo de 2000 ojivas (según los modelos originales de la década del ochenta) necesario para desatar una catástrofe climática global, conocida como *invierno nuclear*. Las consecuencias de un invierno nuclear serían similares a los efectos climáticos que se generarían si impactará un cometa de algunos cientos de km de diámetro sobre la superficie terrestre. La humanidad y gran parte de las especies se extinguirían de la faz de la Tierra de la misma manera que los dinosaurios, hace más de 65 millones de años atrás.

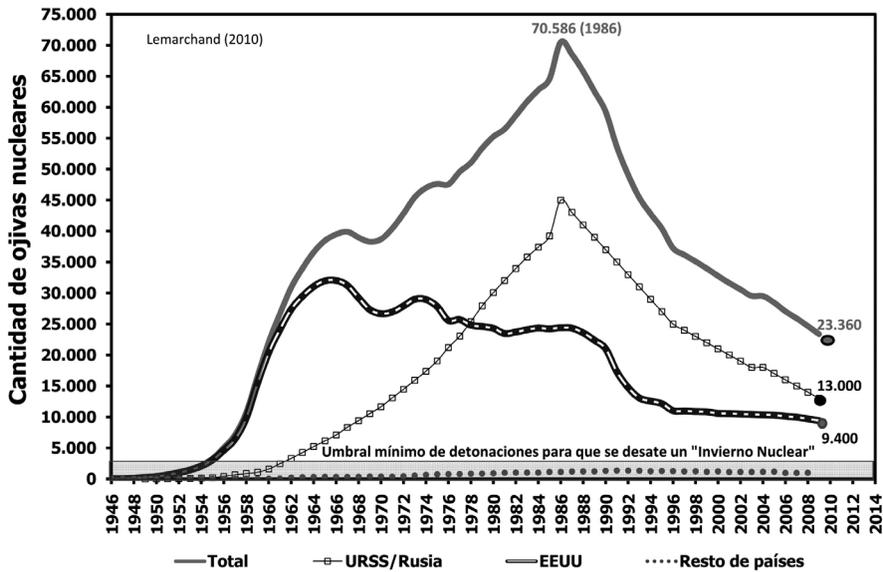

**Fig. 6.** Evolución de la distribución de ojivas nucleares existentes en el planeta (1946-2009). Fuente: Lemarchand (2010)

Recientes estudios desarrollados por Toon et al (2008) y Mills et al (2008) con modelos de la atmósfera terrestre más realistas que los originales del ochenta, demostraron que una pequeña guerra regional que use apenas un centenar de ojivas nucleares, podría generar también una catástrofe global. Asumiendo estos nuevos umbrales, actualmente la humanidad del planeta tiene desplegadas "230 veces más ojivas nucleares" que las necesarias para desatar un invierno nuclear global.

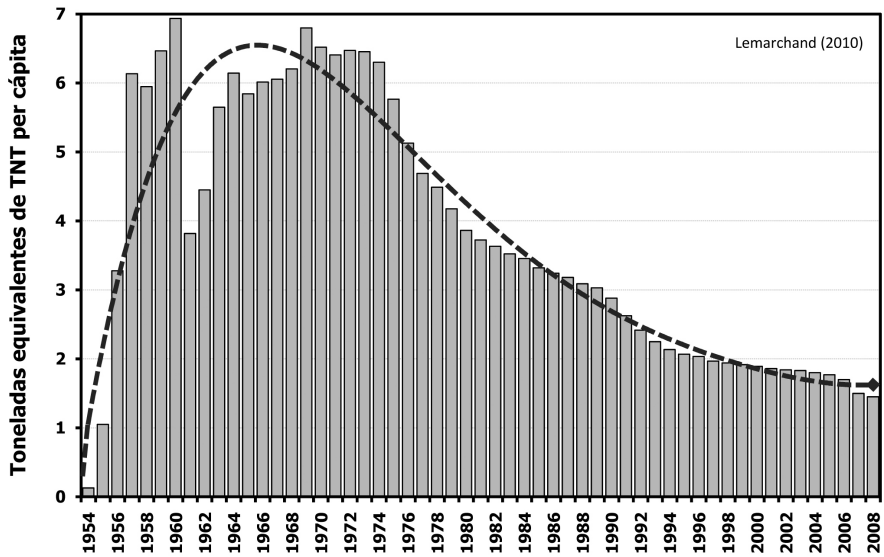

**Fig. 7.** Evolución de la capacidad destructiva equivalente del armamento nuclear desplegado por año dividido el valor la población mundial anual (1954-2008). Fuente: Lemarchand (2010).

Por otra parte, Sullivan et al (1978), mostraron que las transmisiones de radio terrestres han aumentado la intensidad de sus transmisiones de radio entre 1945 y 1980, por lo menos entre 4 y 5 órdenes de magnitud. También mostraron cómo las señales terrestres podrían ser fácilmente detectadas por las eventuales civilizaciones tecnológicas cercanas. Por ejemplo, los Sistemas de Radar de Alerta Temprana, relevan constantemente todo el cielo con potentes transmisores. Las últimas son las más señales, que se escapan al espacio exterior, más intensas generadas por la humanidad, con excepción de las transmisiones de los radares planetarios (por ejemplo, Arecibo, DSN, etc.).

Es interesante observar que nuestra especie alcanzó el nivel de ser detectada por otra civilización tecnológica galáctica al mismo tiempo, que la humanidad alcanzó la capacidad tecnológica de auto-aniquilación.

En trabajos anteriores (Lemarchand 2004, 2006) ha representado la distribución del número de batallas en función de su intensidad (medida como la fracción de muertes generadas sobre el total de participantes). Cuando se analizaron estas distribuciones normalizando el factor de las tecnologías involucradas, se encontró una alta correlación entre el coeficiente de letalidad y la pendiente de la distribución de las fatalidades en las guerras. Un análisis ex-haustivo muestra que estas distribuciones siguen muy bien el comportamiento de los sistemas de criticalidad auto-organizada (SOC).

Los resultados obtenidos siguen leyes de potencia con una pendiente muy similar a la encontrada por Clauset et al. (2007), en sus análisis de la frecuencia de graves atentados terroristas en todo el mundo entre 1968 y 2006. Ambos estudios, se refieren a la dinámica de las muertes generadas por la violencia intra-humana y su combinación contempla desde los asesinatos de unos pocos individuos, a la muerte decenas de millones de personas en las guerras. De alguna manera, esto muestra que la dinámica de la violencia entre los humanos se rige por el mismo tipo de procesos observados en una gran variedad de otros sistemas complejos, que tienen la propiedad de criticalidad auto-organizada (Jensen, 1998). Los estudios sobre la distribución de las guerras ocurridas en los últimos 500 años y su combinación con los datos de los niveles de frecuencia de los ataques terroristas, muestran que la violencia letal entre los seres humanos sigue un comportamiento similar a la dinámica de criticalidad auto-organizada. Con los datos mencionados, el modelo SOC predice la posibilidad de tener un evento en el que toda la población mundial se aniquile dentro de los próximos cientos de años.

Debido a la complejidad, la fiabilidad y el envejecimiento de los sistemas de alerta temprana, las tasas de falsas alarmas de un ataque nuclear han sido siempre relativamente altas. Por lo general, se cree que durante una gran crisis internacional no habría tiempo suficiente para distinguir entre las falsas alarmas y un alerta real de un ataque enemigo. Los datos históricos disponibles sugieren que al menos una falsa alarma lo suficientemente grave como para provocar un ataque nuclear estratégico, sería producida con un 50% de probabilidad durante el tiempo de una crisis internacional larga (Wallace et al. 1986).

Por último, la Figura 8 muestra la evolución anual de los gastos militares mundiales (1950-2008), expresada en miles de millones de dólares constantes de 2009. Para intentar dar significado a esta enorme cantidad de dinero, se la puede comparar con el presupuesto global utilizado anualmente en la exploración espacial (que combina el presupuesto anual de todas las agencias espaciales nacionales e internacionales del mundo). Los gastos espaciales mundiales solo representan menos del 3,5% del gasto militar mundial anual.

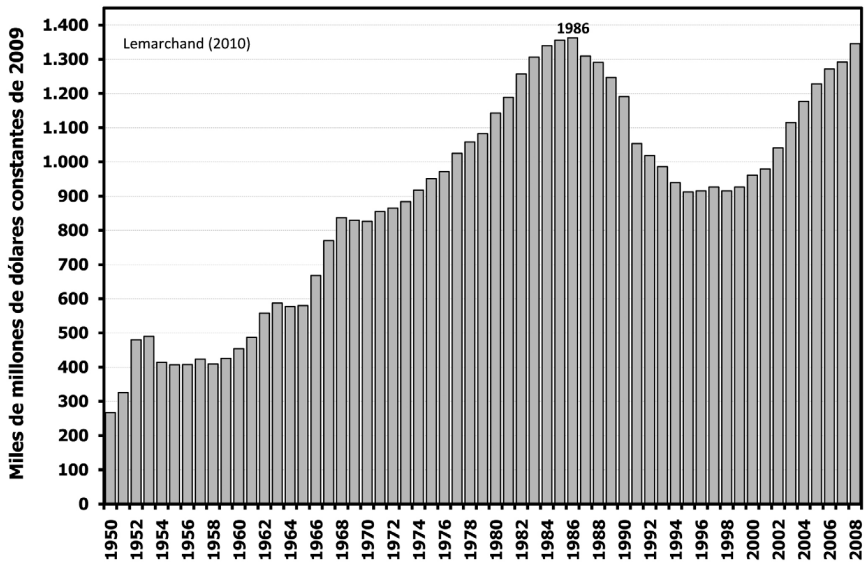

**Fig. 8.** Evolución de los gastos militares mundiales (1950-2008) expresados en dólares constantes del año 2009. Fuente: Lemarchand (2010).

Las Naciones Unidas identificaron las áreas prioritarias mundiales que deben ser atendidas para eliminar del mundo las necesidades básicas insatisfechas, la exclusión social, las desigualdades, el cuidado del medio ambiente y el planeta, la necesidad de garantizar la educación básica, entre otros. El programa establecido por los Estados Miembros se engloba dentro de los Objetivos del Milenio (ODM) de las Naciones Unidas. En la Tabla 3 se compara los gastos militares mundiales con el costo total mundial necesario para cumplir con todos los ODM.

La evolución a largo plazo de los indicadores sociales analizados (Lemarchand 2009), muestran una transición que comenzó tras la Segunda Guerra Mundial y que podría terminar en la segunda mitad del siglo XXI. En una primera aproximación, se estima que este período abarca casi 200 años, con un pico más acentuado entre 1985-2015. Los estudios de la dinámica poblacional y sus modelos (Kapitza 1996) muestran que el mundo está atravesando una transición demográfica similar a la que tuvo hace 12.000 años con la invención de la agricultura y las ciudades. La humanidad también está enfrentando una transición en los procesos democratización mundial, que se inició hace 200 años y persistiría por un lapso de 100 años más (Lemarchand 2006).

**Tabla 3.** El costo de alcanzar los Objetivos de Desarrollo del Milenio de las Naciones Unidas como porcentaje del gasto militar mundial. Fuente: Adaptado de Gillis (2009: 13) usando datos tomados de la gráfica 4 y de la publicación, The Costs of Attaining the Millenium Development Goals, The World Bank, Washington. Accesible en: http://www.worldbank.org/html/extdr/mdgassessment.pdf

| |
|---|
| **Objetivo:** Erradicar la pobreza extrema y el hambre para el 2015 Llevar a la mitad la proporción de personas que viven con menos de 1 dólar diario y sufren hambre **Costo:** USD 39.000 a 54.000 millones **Porcentaje del Gasto Militar Global:** 2,6% a 3,7% |
| **Objetivo:** Promover la educación universal y el equilibrio de género para el 2015 Alcanzar la educación universal y eliminar la disparidad de género en la educación **Costo:** USD 10.000 a 30.000 millones **Porcentaje del Gasto Militar Global:** 0,7% a 2,0% |
| **Objetivo:** Mejorar la salud para el 2015 Reducir en 2/3 la tasa de mortalidad infantil antes de los 5 años, reducir en 3/4 la tasa de mortalidad materna y revertir la difusión del HIV/SIDA **Costo:** USD 20.000 a 25.000 millones **Porcentaje del Gasto Militar Global:** 1,4% a 1,7% |
| **Objetivo:** Medio ambiente sostenible para el 2015 Llevar a la mitad el número de personas sin acceso al agua potable, mejorar las condiciones de vida de más de 100 millones de personas que habitan en villas miserias **Costo:** USD 5.000 a 21.000 millones **Porcentaje del Gasto Militar Global:** 0,3% a 1,4% |
| Los gastos militares globales utilizados solo en el año 2008, equivalen a los costos totales de las Naciones Unidas durante 732 años de funcionamiento o a los de la UNESCO durante 4364 años… |

**Nota de la Tabla 3:** La metodología utilizada por el Banco Mundial para estimar los costos de los Objetivos del Milenio, asume que –debido a la superposición de tareas- resulta muchísimo más económico agrupar distintos Objetivos del Milenio entre si y estimar el costo agregado de lograr dichas metas. Si se optara por estimar el costo en forma individual de cada uno de los 8 objetivos previstos en el programa, la suma total sería muy superior. Por esta razón, en esta tabla se presentan los costos de los objetivos en forma agregada, tal cual se presenta en la estimación del Banco Mundial.

De los datos analizados, se deduce que en términos evolutivos de largo plazo la dinámica poblacional humana (desde el *Homo sapiens* en adelante) muestra un crecimiento auto-similar siguiendo una ley de potencia. El modelo matemático desarrollado por Kapitza (1996) explica a la "transición demográfica actual" como una propiedad intrínseca del sistema. En este período de transición, toda perturbación (social, económica, militar, etc.), aun pequeña, será amplificada por el propio sistema a escala global. Esto determina un período en donde se favorece la *oportunidad* de que un pequeño evento violento adquiera rápidamente un alcance mundial.

Los estudios de distribución de muertes generadas por hechos violentos muestran, con datos que se extienden a lo largo de 7 órdenes de magnitud (desde las muertes individuales por ataques terroristas al número de fallecidos durante la Segunda Guerra Mundial) que la violencia intra-humana sigue una dinámica de criticalidad auto-organizada. Esta permite estimar la *probabilidad* de que ocurra un evento violento en el que toda la especie se auto-aniquile. Un fenómeno de estas características podría ocurrir durante los próximos siglos.

La tasa de crecimiento exponencial en el coeficiente de letalidad del armamento disponible, permite desde finales de la Segunda Guerra Mundial, que la humanidad disponga de la *tecnología necesaria* para asegurar su total extinción.

Un somero análisis de la inversión anual en gastos militares muestra el *excesivo financiamiento* que cuentan estas actividades, en comparación con las que reciben otras necesidades prioritarias de la humanidad.

Por lo tanto, se tiene la *oportunidad*, la *probabilidad,* la *tecnología* y el *financiamiento* para que ocurra un evento en el cual la humanidad se puede extinguir. Para simplificar el estudio no se ha mencionado en este análisis las consecuencias de la degradación ambiental, el cambio climático, las epidemias, etc.

La distribución de las series temporales analizadas, muestra que esta *Edad de Adolescencia Tecnológica* humana, se extiende por un período de $L_{Min}$ = 200 años. Asumiendo que lo que ocurre aquí en la Tierra es esencialmente la media de lo que ocurre en el resto del universo (Principio de Mediocridad), se puede tomar este valor como un umbral mínimo.

## 5. El impacto del valor de vida media de las civilizaciones en el diseño de la búsqueda de señales artificiales de origen extraterrestre.

La Ecuación de Drake (Pearman, 1963) asume que el número de civilizaciones tecnológicas en la galaxia está representado por la multiplicación de una serie de factores astronómicos, biológicos y societales. El propio Drake considera que con buena aproximación la ecuación se puede reducir a N $\simeq$ 1,5 x $L$. En el caso analizado en la Sección 4, $L_{Min}$ = 200 años. Por lo tanto, el mínimo número de civilizaciones tecnológicas presentes en la galaxia será del orden de $N_{Min}$ = 300.

Suponiendo que todas las civilizaciones tecnológicas $N$, residen en un disco galáctico de radio $R_g$, y de espesor de $H_g$, la densidad galáctica de civilizaciones será descripta por la siguiente ecuación (Lemarchand 2000a):

$$\delta(N) = \frac{N}{\pi \times H_g \times R_g^2} = \frac{300}{\pi \times 0,6 \times 15^2} \text{ civilizaciones} \times \text{kpc}^{-3} \cong 0,7 \text{ civilizaciones} \times \text{kpc}^{-3}$$

Aquí se ha asumido un valor de $H_g$ = 0,6 kpc y $R_g$ = 15 kpc. El último resultado propone un umbral mínimo de una civilización galáctica por cada 1,42 kpc³. Un valor $L_{min}$ ~ 200 años también implica que otras civilizaciones no deberían tener tiempo suficiente para desarrollar transmisores muy potentes, por lo que implícitamente se está infiriendo que este tipo de civilizaciones tendrían un nivel tecnológico similar al terrestre. Si estas civilizaciones galácticas estuvieran distribuidas de manera uniforme, pero al azar, la probabilidad de encontrar otra civilización galáctica en una esfera de radio $r$ alrededor de la Tierra estará dada por:

$$P = 1 - e^{-\delta(N) \times \frac{4}{3} \times \pi \times r^3}$$

La mayor potencia equivalente de radiación isotrópica (EIRP) que dispone el planeta lo tiene el radar de Arecibo, en Puerto Rico (EIRP$_{Arecibo}$ ~ 2 x 10$^{13}$ W). Se podría razonablemente suponer la existencia de una civilización galáctica de características similares a la terrestre que utilice transmisores de EIRP$_{ET}$ ~ 10$^{14}$ W (la humanidad ya tiene la capacidad tecnológica como para construir este tipo de transmisores). Teniendo en cuenta la sensibilidad actual de los equipos empleados en SETI, una observación desde el observatorio de Arecibo sería capaz de detectar una señal de $10^{14}$ W a una distancia $r \sim 1.4$ kpc.

Si se analiza el caso del observatorio ATA (350 antenas de 6 m cada una), éste sería capaz de detectar la misma señal desde una distancia $r \sim 0.35$ kpc. Si se asume que el observatorio terrestre se encuentra apuntando en la dirección correcta en el momento correcto (sincronización perfecta entre transmisión y recepción), las observaciones realizadas desde Arecibo tendrán una probabilidad P ~ 0.8 de detectar dichas señales, mientras que las posibilidades de ATA –ante las mismas condiciones de observación-  será solo de P ~ 0,02.

Está claro que la hipótesis de una estrategia transmisión omnidireccional no es sustentable en el largo plazo. Por eso se hace imprescindible buscar nuevas estrategias para coordinar los regímenes de coordinación *a priori* entre las "transmisiones" y "recepciones" (Lemarchand, 1994).

Sólo las civilizaciones galácticas ubicadas a distancias < 0,02 kpc (~ 65 años luz) tienen la posibilidad de detectar señales de radio artificiales procedentes de la Tierra, y por lo tanto, en el presente, la humanidad solo tiene la capacidad de detectar hipotéticas "respuestas" de aquellas estrellas que se encuentran a distancias < 0,01 kpc. También se podría estar recibiendo señales de las civilizaciones tipo terrestres, que ya hubieran detectado planetas del tamaño de la Tierra, con instrumentos similares a la Misión Kepler, u otros observatorios espaciales que se encuentran en estado de construcción.

Si nuestra civilización terrestre es una civilización "típica" en la galaxia (Principio de Mediocridad) y si todas las civilizaciones pasan por la misma *Edad de Adolescencia Tecnológica*, al igual que la Tierra en el presente, entonces resulta lícito esperar que no existan simultáneamente más de 300 civilizaciones con capacidad comunicativa en la galaxia. Este hecho estaría de acuerdo con el resultado nulo, obtenido hasta el momento, durante más de 50 años de observaciones.

En 1959, Cocconi y Morrison finalizaban su artículo afirmando: "…la probabilidad de éxito de este tipo de investigación es muy reducida.  Sin embargo, si nunca se lleva a cabo, ésta siempre será cero." Medio siglo después, esta sentencia sigue teniendo una parsimoniosa validez.

## 6.    Ejercicios:

- Utilizando los valores de potencia de transmisión y sensibilidad del observatorio de Arecibo, determine cuál será la distancia máxima desde la que se podría detectar una señal emitida desde una antena transmisora idéntica a la de Arecibo en la frecuencia de 6 GHz. Asuma una relación señal-ruido de SNR=3. Puede encontrar los detalles técnicos del observatorio en internet.
- Calcule cuál sería la probabilidad de detección de una señal de las características anteriores, si existieran 100.000 civilizaciones tecnológicas transmitiendo mensajes del descripto en el ejercicio anterior. Asuma una distribución homogénea de civilizacionesen la galaxia.

## Agradecimientos

## Referencia